\documentclass{article}
\usepackage[T1]{fontenc} 
\usepackage[utf8]{inputenc} 
\usepackage{ismir,amsmath,amssymb,cite,url}
\usepackage{graphicx}
\usepackage{color}
\usepackage{float}
\usepackage{siunitx}

\newcommand*\rot{\rotatebox{90}}
\usepackage{xcolor}
\newcommand{\new}[1]{{#1}}

\title{TheGlueNote: learned representations \\ for robust and flexible note alignment}

\oneauthor
  {Silvan David Peter$^1$ \hspace{1.5cm} Gerhard Widmer$^{1,2}$} {$^1$ Institute of Computational Perception, Johannes Kepler University Linz, Austria \\ $^2$ LIT AI Lab, Linz Institute of Technology, Austria}

\def\authorname{S. Peter and G. Widmer}

\begin{document}

\maketitle
\begin{abstract}
Note alignment refers to the task of matching individual notes of two versions of the same symbolically encoded piece.
Methods addressing this task commonly rely on sequence alignment algorithms such as Hidden Markov Models or Dynamic Time Warping (DTW) applied directly to note or onset sequences.
While successful in many cases, such methods struggle with large mismatches between the versions.
In this work, we learn note-wise representations from data augmented with various complex mismatch cases, e.g. repeats, skips, block insertions, and long trills.
At the heart of our approach lies a \new{transformer encoder} network --- TheGlueNote\footnote{\new{\url{https://github.com/sildater/thegluenote}}} --- which predicts pairwise note similarities for two 512 note subsequences.
We postprocess the predicted similarities using flavors of weightedDTW and pitch-separated onsetDTW to retrieve note matches for two sequences of arbitrary length.
Our approach performs on par with the state of the art in terms of note alignment accuracy, is considerably more robust to version mismatches, and works directly on any pair of MIDI files.

\end{abstract}

\section{Introduction}\label{sec:introduction}

Note alignment refers to the task of matching individual symbolically encoded notes in two versions of the same piece.
Note matches can be derived for any two versions, however, this task is usually addressed for pairs of MIDI performances and scores encoded in various formats.
The resulting performance-to-score alignments provide the data for several research directions in MIR and computational musicology, such as expressive performance generation, score quantization, and performance research.

To match notes can sometimes be a near-trivial task, especially with well corresponding versions, minimal expressive playing, and simple pieces.
However, more often than not the unaligned data of interest and availability does not fit these criteria:
performers make mistakes; play extra repeats, variations, and ornamentations; rehearsal recordings discontinue or restart; automatic transcriptions contain various amounts of note mismatches; and the musical material tends towards the virtuosic, dense, and complex.

Due to its close similarity with sequence alignment, note alignment is usually approached with flavors of Dynamic Time Warping (DTW) or Hidden Markov Models (HMM) based on note or chord representations.
Such representations are typically localized, and the alignment methods process them sequentially. 
The aforementioned common difficulties in MIDI performances do not harmonize well with these constraints:
e.g., differently ordered chord onsets clash with DTW's monotonicity condition,
trills create (sometimes substantial) mismatches with similar pitches and thus misleading local distances,
and repeats, skips, recording takes, etc.~introduce large mismatches between the sequences which require a more zoomed-out perspective.
To be clear, there is nothing that a priori prevents sequence alignment methods from working in these scenarios, however, in practice, the propensity of alignment methods for propagating errors render the matching quality hit-and-miss.

In this work, we address note alignment via learned representations which leverage non-local information, \new{i.e., the entire sequence of notes influences the representation of each note}.
We train an attention-based encoder --- TheGlueNote --- to predict note representations for two 512 note subsequences.
Before being passed to the network, the subsequences are augmented with a variety of challenging and large mismatch cases.
At the network's output, we compute a pairwise similarity matrix between the note representations and compare this matrix to target note matches via two classification loss terms.
That is, the notes are guided towards similar representations if they match, and dissimilar representations for all others.
In the process, TheGlueNote is trained to robustly predict note similarities even in the presence of substantial mismatches.

We took care to design TheGlueNote as annotation-agnostic as possible.
Prior approaches mitigate edge cases by introducing additional submethods, e.g., by modeling left-right hand streams separately, excluding notated ornaments from certain steps, or requiring coinciding chord notes (see section~\ref{sec:related_work}).
This introduces limitations on the types of files which can be processed, requiring staff or voice information, scores with ornament information, or even just quantized scores.
In contrast, our model is trained directly on data from MIDI files with no quantization or score annotation requirement.

To extract final note matches from the similarity matrix, we present three possible additional methods.
First, we simply match the notes with maximal similarity.
Second, we add a decoder head to our note representation backbone -- a network which predicts matching notes based on the similarity matrix.
Third, we use DTW alignment techniques to extract a mapping from the similarity matrix and in turn use this mapping to match individual notes.
Putting the pieces together, TheGlueNote leverages learned representations for note alignment, performs on par with the state of the art, excels at complex mismatch cases, and works with plain MIDI input data.

The rest of the paper is structured as follows:
Section~\ref{sec:related_work} discusses related work.
Section~\ref{sec:model} describes the model architecture and match extractor variants.
Section~\ref{sec:experiments} introduces training specifications, data processing, and metrics which are applied in Section~\ref{sec:evaluation} where we evaluate the trained model in an ablation study on the variations of architecture and match extraction, and in comparison with state-of-the-art methods both in regular and complex mismatch cases.
Section~\ref{sec:discussion}, the discussion, concludes the paper.

\section{Related Work}\label{sec:related_work}

Note alignment is a basic technology vital to many downstream tasks in symbolic music processing and computational musicology.
We structure this review of related literature into a part on current state-of-the-art methods, relevant work in the neighboring domains of audio and real-time alignment, and pertinent literature on matching tasks for non-music data. 

Note alignment methods almost universally make use of either Dynamic Time Warping or Bayesian Networks on pitch-based representations of either individual notes or chords~\cite{gingras2011improved,chen2014improved, Nakamura2014MergedOutputHM,Nakamura:2015,nakamura2014outer,nakamura2017performance,nakamura2015autoregressive, peter2023tismir,peter23offline}.
As the principal formulation is straightforward, most recent efforts have focused on formalizations and heuristics that mitigate specific problems in edge cases. 
Skips and repeats present a major difficulty which can be directly modelled at the cost of computational complexity~\cite{nakamura2014outer} or side-stepped if the use of annotated anchor points is possible~\cite{peter2023tismir}.
Another difficulty are ornamentation notes which can be modelled as separate states~\cite{Nakamura:2015} or excluded from a first coarse alignment and handled separately in a fine-grained note matching step~\cite{peter23offline}.
Nakamura et al.~\cite{Nakamura2014MergedOutputHM} further model left-right hand asynchrony in piano performance.
In their most recent work, note alignment is framed as a hierarchical refinement with explicit modelling of an alignment error identification step~\cite{nakamura2017performance}.
In recent work by Peter~\cite{peter23offline}, sequence non-ordinality is mitigated by a score-based chord representation, the resulting model is thus limited to performance to score alignment.
The current state of the art (SOTA) which we will use for reference in this work is given by two DTW-based methods \cite{peter23offline,peter2023tismir} and one HMM-based method \cite{nakamura2017performance}.

Realtime alignment or score following methods have been developed since the 1980s~\cite{dannenberg1984line,vercoe1984synthetic} and largely mirror the previous methods in their core elements: On-Line Time Warping (OLTW),~\cite{cancino2023accompanion} and Bayesian Networks, in particular HMM~\cite{Nakamura2014MergedOutputHM, Raphael2009OrchestralAF, nakamura2015autoregressive}.

Alignment of musical audio is an important idea generator for symbolic note alignment.
For introductions of audio alignment, we refer the reader to Arzt~\cite{arzt:2016} and Müller~\cite{Mueller:2015}.
Audio alignment is prone to memory and compute bottlenecks.
Several versions of DTW addressing these issues have been developed~\cite{muller2021sync}. 

Deep Learning has largely been absent from music alignment, with notable exceptions in real-time audio-image matching~\cite{matthias_dorfer_2018_1492535,henkel2019score} and in symbolic score following~\cite{peter23offline}.
On the other hand, we take inspiration from image processing, in particular from the task of local feature matching~\cite{xu2024local}: the matching of pixels encoding the same location on an object in two images of the same object.
Local feature matching often uses neural network-based feature extractors~\cite{lindenberger2023lightglue,sarlin2020superglue} and our proposed model in particular is informed by the MatchFormer~\cite{sarlin2020superglue}.

\begin{figure*}[h]
 \centerline{
 \includegraphics[alt={A flowchart of the proposed models. The center row shows the main model TheGlueNote, a transformer encoder with all its modules, and a Decoder Head, another transformer encoder.
 The top row includes modules used during training, starting the synthetic data augmentation from a single MIDI file, and ending with three possible loss terms.
 The bottom row shows modules used during inference, starting from two MIDI files, and ending with three possible match extractors.},width=1.0\linewidth]{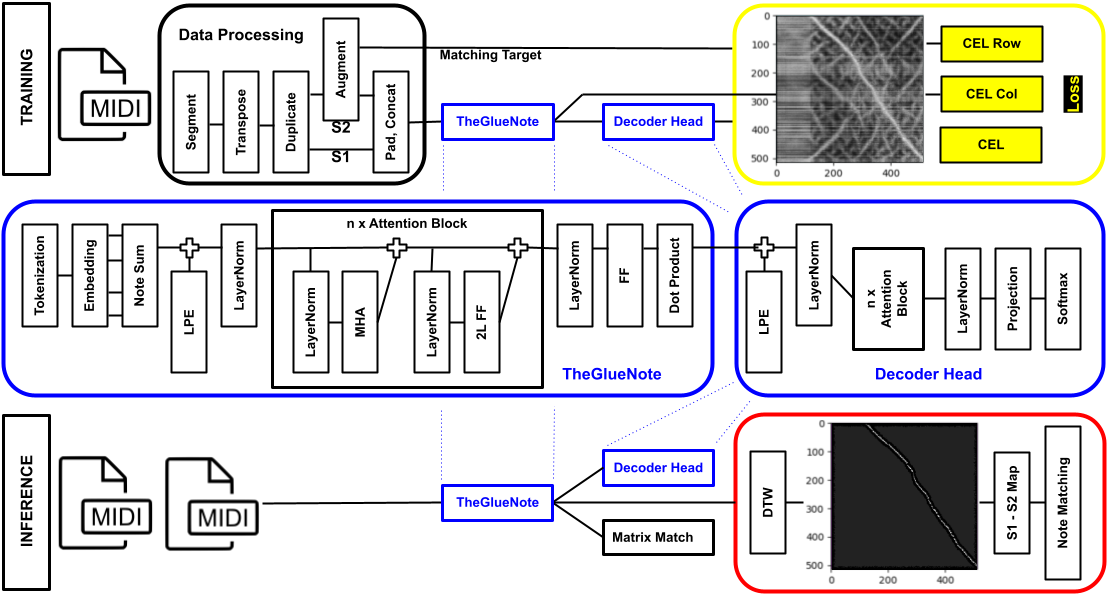}}
 \caption{Overview of the proposed model. 
 During \textbf{training} (top row), the data flows from Data Processing (black, top left) through TheGlueNote (blue, middle left) into the Decoder Head (blue, middle right) and the aggregated Loss (yellow, top right).
 Concretely, a MIDI file is loaded into the Data Processing module which outputs matching targets to the loss module and 
 the concatenated sequences $s_1$ and $s_2$ to TheGlueNote.
 TheGlueNote (middle row) consist of a transformer encoder with learned positional embeddings (LPE) and repeated attention blocks (center module with multihead-attention MHA and a two-layer feedforward network 2L FF).
 The note-wise representations are split and multiplied for a pairwise similarity matrix with $s_1$ in the row and $s_2$ in the column dimension shown in the loss module.
 Two cross-entropy loss terms are computed from this matrix and it is also forwarded to the decoder head whose classifier output adds a third loss term.
 During \textbf{inference} (bottom row), two MIDI files to be matched are directly passed to TheGlueNote.
 The resulting similarity matrix can be processed in three ways: 1) direct maximal similarity match extraction (Matrix Match box) 2) using the decoder head's output, or 3) using a DTW-based match extraction (red, bottom right).
 }
 \label{fig:flowchart}
\end{figure*}

\section{Model}\label{sec:model}

In this article, we present a model which is trained to create note representations for two sequences such that the representations' pairwise similarity corresponds to the sequences' alignment ground truth.
Using these representations, we aim to uniquely match individual notes.
The proposed model consists of a fixed-length tokenization, an encoder backbone, and a dual classification loss.
Furthermore, we introduce three variants of match extraction from the model's output similarity matrix: direct similarity matrix processing, using a decoder head for classification, and DTW-based match extraction. \figref{fig:flowchart} presents an overview of the components.

\subsection{TheGlueNote}\label{subsec:thegluenote}
At the heart of our model is a non-causal transformer encoder (see \figref{fig:flowchart} middle left).
Its purpose is to learn note representations for two note sequences $s_1$ and $s_2$, and its target is the alignment between the sequences.
A pairwise similarity matrix computed between the note representations of two sequences mediates between output and target. 
We treat the matrix as a match classifier for each note, i.e., for each row (a note in $s_1$) the column (a note in $s_2$) with maximal similarity should correspond to matching notes, and vice versa. 

Technically, two at least partially matching subsequences $s_1$ and $s_2$ of 512 notes each 
are prepended with a default note and processed using the fixed-length structured tokenization~\cite{hadjeres2021piano,miditok2021}, which encodes relative onset, pitch, duration, and velocity.
The now 513 note (2052 token) sequences $s_1$ and $s_2$ are concatenated and passed to the encoder.
The encoder sums the four tokens per note and adds a learned positional embedding for a note-wise sequence of length 1026.
Layer normalization is applied before the first encoder block and within the attention and feedforward blocks but not again on the residual stream.
Self-attention is applied to the full concatenated sequence, which amounts to combined within-sequence attention ($s_1$-$s_1$, $s_2$-$s_2$) and between-sequence attention ($s_1$-$s_2$, $s_2$-$s_1$).
The final residual is normalized after the last block and fed through a single dimension-conserving linear layer.
For dimensions and hyperparameters of different versions see section \ref{subsec:training} and in particular \tabref{tab:hparams}.

The 1026 final output vectors of size corresponding to the residual stream are treated as representations of individual notes. 
The sequences are split again and pairwise similarities between all 513 notes of $s_1$ and all 513 notes of $s_2$ are obtained via dot product. 
The resulting similarity matrix ($s_1$ in the rows and $s_2$ in the columns, see \figref{fig:flowchart} top right) is compared to two classification 
targets: Softmax across row dimension is the model's prediction of the matching note in $s_1$ for each note in $s_2$ (except for the prepended default note), and softmax across the column dimension  is the model's prediction of the matching note in $s_2$ for each note in $s_1$ (again, except for its prepended default note). 
Both are compared against the ground truth via a cross-entropy loss (CEL).
Unmatched notes in the ground truth receive a target corresponding to the default note in the other sequence, the default notes itself receive no loss.

\subsection{Match Extractors}\label{subsec:matchextracor}

Note similarities are a useful intermediary, however, they do not yet define note matches.
In this section, we detail three possible note match extractors.
The simplest way of producing matches is to directly use \textbf{similarity matrix-based match extraction}.
That is, for each note in $s_2$, we match it to its most similar note in $s_1$, including the default note (=unmatched) as possibility.
A little bit of index housekeeping avoids conflicting matches and notes without prediction.

A second approach is to train TheGlueNote with an additional \textbf{decoder head for match extraction}.
The decoder is also an non-causal neural network with the same high-level structure as the encoder\footnote{The "decoder head" is technically also a transformer encoder without memory input, however, it decodes the representation towards classification logits, so we opt for this name.} (see \figref{fig:flowchart} middle right).
The decoder head processes the pairwise similarity matrix for each actual note of $s_2$ (hence a 512 by 513 matrix, excluding the default note in $s_2$) and directly predicts the matching note in $s_1$ via 513 output logits (including the default note in $s_1$ for unmatched notes).
During training, its classification CEL is added to the other two losses in an unweighted fashion (see \figref{fig:flowchart} top right).

\textbf{DTW match extraction} offers a way of introducing meaningful constraints to the note matching.
Concretely, naive note matching via maximum similarity in the prediction treats every note separately and ignores information on predictions for notes in its vicinity and previously matched notes.
To introduce this information, we adapt the DTW-based mapping and matching procedure introduced by Peter~\cite{peter23offline} for similarity matrix post-processing.

DTW extraction is split in two processes (see \figref{fig:flowchart} bottom right).
First, understanding similarity as the reciprocal of learned pairwise distance, we use the network's output as input to a weighted DTW with possible directions $[[0,1],[1,1],[1, 0]]$ and associated weights $[1,2,1]$.
This choice of weights normalizes the directions under the Manhattan distance, i.e., any direction is equally costly overall and the diagonal is not favored.
A standard DTW path is computed starting at the top left and ending at the bottom right of the similarity matrix.
Note that extracting a minimizing path through the learned distances discards information relevant to local non-ordinality in favor of added information about each note's neighborhood.
The extracted path should not be used as direct note match prediction, instead it defines a coarse sequence to sequence mapping $m:\mathbb{R} \rightarrow \mathbb{R}$ by linear interpolation between the onset times of notes in the path.

In a second process, we separate all notes in both sequences by pitch.
For subsequences $s_{1}^{p}$ and $s_{2}^{p}$ of pitch p, we match onset sequences using a DTW pass to find onset pairs that minimize the distance between $s_{2}^{p}$ and $m(s_{1}^{p})$.
Newly matched notes are then added to or overwritten in the original DTW path and the mapping $m$ is updated.
If the MIDI files to be aligned do not fit within the 512 note contexts of the model, which is often the case, we compute several similarity matrices for 512 note windows with a stride of 256 notes. 
We then aggregate the resulting output matrices to a global similarity matrix.

\begin{table}[H]
 \centering
 \begin{tabular}{|l|l|}
  \hline
  Feature & Noise and Mismatch Probabilities\\
   \hline
Tempo $T_{t}$ & $gT_{t}2^{n_{t}}, g \sim  \mathcal{N}(1,0.5), n_{t} \sim  \mathcal{N}(0,0.5)$\\ 
Onset $O_{t}$ & $O_{t}+ n_{t}, n_{t} \sim  \mathcal{U}(-50,50)$\\
Velocity $V_{t}$ & $V_{t}+ n_{t}, n_{t} \sim  \mathcal{U}(-10,10)$\\    
Duration $D_{t}$ & $D_{t}+ n_{t}, n_{t} \sim  \mathcal{U}(-250,250)$\\       
    \hline
Repeats & $\mathcal{P}_{repeat} = 1, \#note_{repeat} \sim \mathcal{U}(8,200)$\\
Skips & $\mathcal{P}_{skip} = 1, \#note_{skip} \sim \mathcal{U}(8,200)$\\
Insertions & $\mathcal{P}_{insertion} = 0.2$, random location\\
Deletions & $\mathcal{P}_{deletions} = 0.2$, random location\\
Trills & $\mathcal{P}_{trill} = 1, \#note_{trill} \sim \mathcal{U}(20,100)$\\
  \hline
 \end{tabular}
 \caption{Augmentations to synthesize complex mismatch cases.
 Four noise terms are added to note features in the first row terms. 
 Sampled noise is clipped to avoid degenerate cases like negative durations. 
 Duration and onset noise are indicated in MIDI ticks.
 Skips, repeats, and trills are introduced with the indicated probability and uniformly sampled length.
 Insertions and deletions are added at random locations with overall probabilities given.}
 \label{tab:aug}
\end{table}

\section{Experiments}\label{sec:experiments}

We report several experiments to asses the qualities of our proposed model. 
In this section, we describe the dataset, data preprocessing, and training as well as model configurations.

\begin{table*}[h]
 \centering
\begin{tabular}{|l|rrr|rrr|rrr|rrr|}
\hline
Data Source & \multicolumn{9}{c|}{Vienna 4x22} & \multicolumn{3}{c|}{Training Data} \\
\hline
Match Extractor & \multicolumn{3}{c|}{Sim Matrix}  & \multicolumn{3}{c|}{Decoder Head} & \multicolumn{3}{c|}{DTW} & \multicolumn{3}{c|}{n.a.}\\
\hline
Unit & Prec & Rec & F  & Prec & Rec & F  & Prec & Rec & F & TL & VL & VA \\
\hline
Pitch-Onset Similarity Matrix &    7  &       7  &       7  &      -  &      -  &      -  &      85  &      89  &      82 &  - &   - &   - \\
\hline
TGN-large &    97  &      97  &      97  &      96  &      97  &      96  &      99  &     100  &      99 &    0.183 &    0.126 &   0.958 \\
TGN-mid &   81  &      81  &      81  &      87  &      88  &      87  &      99  &      99  &      98  & 0.171 & 0.145 & 0.996\\
TGN-small &   75  &      75  &      75  &      83  &      87  &      81  &      99  &      99  &      99  &    0.374 &    0.280 &   0.902 \\
\hline
\end{tabular}

 \caption{Ablation of Model configuration and match extraction. 
 All match results are computed on the Vienna4x22 Dataset.
 Values reported are average note match precision, recall, and F-score across all performances.
 The results are computed for each match extractor / model combination.
 The first line serves as a simple baseline for the similarity matrix-based and the DTW-based match extractors:
 we report results of their processing a pitch and onset-based similarity matrix.
 We further report the average training loss (TL), validation loss (VL), and validation accuracy (VA) in the final epoch for each model.}
 \label{tab:ablation}
\end{table*}

\subsection{Data}\label{sec:data}
A data sample for our model is a pair of 512 note subsequences.
Note alignment ground truth data of real pieces and performances is available~\cite{peter2023tismir,vienna4x22,hu2023batik}, however, this data is biased towards cases where prior note alignment methods could successfully be applied.
To present the model with a wide variety of (mis)match cases we use synthetically augmented MIDI data for training.
The original MIDI tracks are taken from the (n)ASAP dataset~\cite{peter2023tismir}, albeit not its note alignments, only the score and performance MIDI files directly.

Ground truth match data is created entirely synthetically by copying each MIDI file and augmenting it with a combination of the processes which we describe in the following, and whose parameters are given in \tabref{tab:aug}.
\new{The original inter-onset intervals (as a proxy for tempo) are stretched by a global factor $g$, and by note-wise factors $n_t$, these factors are multiplicative and normally distributed.
Note onsets and durations are also changed note-wise, yet by additive uniform noise in MIDI ticks.
All MIDI files are encoded using 480 ticks per beat and 120 beats per minute, one MIDI tick is thus slightly longer than one millisecond.
Velocities are modified by additive uniform noise within the 128 standard MIDI velocity values.
For repeats, skips, and trills, the probability of generating the mismatch per 512 note sequence is given, as well as note quantities sampled uniformly.
The mismatches are inserted contiguously into the sequence and there is at most one augmentation of each mismatch type per 512 note sequence.
Finally, insertions and deletions are generated from the existing notes, each note is deleted or copied and randomized (i.e., inserted) with the given probability.
All augmentations are recomputed for every batch of training data.
The values in \tabref{tab:aug} are given for reproducibility and transparency, although different variations were tested, we do not claim that these are optimal values.}

We further add transposition to the maximal extent available on a piano keyboard as general augmentation affecting both subsequences.
The augmentation is carried out in the data loader so each epoch will produce different samples from the 1032 valid MIDI files in our dataset.
Data augmentation is only used during training, at inference two MIDI files are matched as is (see \figref{fig:flowchart}).

For testing, we obtained the exact test files used in the reference literature~\cite{peter23offline}. 
These files stem from proprietary datasets~\cite{CancinoChacon:2017ht,magaloff} and were chosen due to the alignment complexity they provide.
To test for robustness under challenging mismatch scenarios, we augment these performances for an experiment including extended (100+ note) mismatches that cover approximately 20\% of the notes.
\new{Each 512 note subsequence pair contains exactly two mismatching segments, one in $s_1$ and one in $s_2$, each segment is contiguous and its notes are randomly sampled.
Note that such randomized contiguous mismatches are different from the synthetic mismatch segments seen during training (i.e., trills, repeats, skips, see \tabref{tab:aug}).}
For comparison with our reference models, we have to limit ourselves to score to performance alignment instead of general MIDI to MIDI alignment, as some of the compared models only work with this type of musical material.
To compare different model configurations and match extractors, we further use the public Vienna 4x22 Dataset~\cite{vienna4x22}.
This dataset consists of 4 pieces with 22 performances each.
The pieces are comparatively simple and mismatches minimal.

\begin{table}[H]
 \begin{center}
 \begin{tabular}{|l|r|c|c|c|c|r|}
  \hline
  Model & \#p & rd & \#b & \#h & bs & \#ph \\
   \hline
  TGN-large & 28M & 512 & 8 & 8 & 8 &  27M \\
  TGN-mid & 5.7M & 256 & 6 & 8 & 16 &  2M \\
  TGN-small & 1.1M & 128 & 4 & 8 & 24 &  0.6M \\
    \hline
 \end{tabular}
\end{center}
 \caption{Hyperparameters for TheGlueNote (TGN) variations: \#p = parameter count, rd = residual dimension, \#b = number of blocks, \#h = number of attention heads, bs = batch size, and \#p dh = parameter count of decoder head. Parameter counts of the TheGlueNote models (\#p) and their decoder heads (\#ph) are approximate.}
 \label{tab:hparams}
\end{table}

\subsection{Training Setup}\label{subsec:training}

We train model variations differentiated in three sizes.
All our models are trained on a single GeForce GTX 1080 Ti with 12 GB of memory.
We train for 200k steps, independent of batch size, which is set to the maximal capacity of the GPU for each model.
The learning rate is initialized at $5*10^{-4}$ and is scheduled using cosine annealing with warm restarts at an interval of 2k steps.
\tabref{tab:hparams} details the hyperparameters for model variations.
For all attention blocks, the inverted bottleneck of the feedforward network is four times the residual dimension.

\section{Evaluation}\label{sec:evaluation}

In this section, we evaluate our proposed model.
The first part compares different model configurations, the second part compares against state-of-the-art reference methods. 
To evaluate our models, we use note matching precision, recall, and F-score as our main metrics. 
We further report mean final classification losses of the predicted similarity matrix which corresponds to direct note matching on the training and validation data as well as the runtime of different model setups.

\subsection{Ablation Study of Model Configurations}

In a first experiment we train three model configurations.
We evaluate their note matching quality on the Vienna4x22 dataset using three different match extractors:
direct similarity matrix processing, decoder head prediction, and DTW-based match extraction.
\tabref{tab:ablation} details the results.
For all model configurations the match extractors are clearly ranked with DTW-based processing the most promising.
DTW-based match extraction in itself is, however, not enough for good matching.
To illustrate this point, we compute a simple pitch and onset based similarity matrix (the closer in pitch and onset, the higher the similarity) akin to what would be used to assess local distances in standard approaches.
We then apply the similarity matrix-based match extractor and the DTW-based match extractor directly on this matrix.
The first row in \tabref{tab:ablation} shows these reference methods, which perform subpar.

\begin{table*}[h]
 \centering
\begin{tabular}{|l|rrrrrr|r|rrrrrr|r|}
\hline
Data Source & \multicolumn{7}{c|}{Default Data} & \multicolumn{7}{c|}{ 20 \% Mismatch Data} \\
\hline
Piece & \rot{B. Op.~53 3rd.~m.} & \rot{C. Op.~9 No.~1}  & \rot{C. Op.~9 No.~2} & \rot{C. Op.~10 No.~11} & \rot{C. Op.~60} & \rot{mean of 5 pieces} & \rot{total runtime} & \rot{B. Op.~53 3rd.~m.} & \rot{C. Op.~9 No.~1}  & \rot{C. Op.~9 No.~2} & \rot{C. Op.~10 No.~11} & \rot{C. Op.~60}  & \rot{mean of 5 pieces} &  \rot{total runtime} \\
\hline
Unit & \multicolumn{6}{c|}{Match F-Score in \%} &  \si{\second} & \multicolumn{6}{c|}{Match F-Score in \%} &  \si{\second} \\
\hline
Nakamura HMM &    98  &      99  &      98  &      94  &      95  & 98  &     152  &      39  &      65  &      35  &      20  &      63  & 44  &    6458 \\
Peter AutomaticNoteMatcher  &    99  &      84  &      94  &      96  &      89  & 92 &     588  &      82  &      74  &      89  &      71  &      75  &  78 &    808 \\
Peter DualDTWMatcher &     99  &      98  &      99  &      96  &      98  &  98 &    96  &      85  &      96  &      94  &      80  &      83  &  88 &   208 \\
\hline
 TGN-large + DTW   &  99  &      99  &      98  &      96  &      97  & 98 &     33  &      94  &      96  &      95  &      93  &      94  &  94 &    42 \\
 TGN-mid + DTW &    96  &      98  &      98  &      96  &      98  &  97 &     27  &      92  &      95  &      96  &      92  &      95  & 94 &     38 \\
TGN-small + DTW &      99  &      98  &      98  &      96  &      97  &  98 &    21  &      94  &      97  &      95  &      93  &      94  & 95 &     31 \\
\hline
\end{tabular}
 \caption{F-scores of three reference models and our proposed models across five challenging pieces.
 The matching results are given as f-scores in \% and the runtime in seconds.
 The data is split in two groups: a default case with the original performances, and a mismatch case, where challenging skips and repeats which in total constitue \~20\% of the notes have been introduced.
 The models are split in two groups: three state-of-the-art reference models and our proposed model in three configurations.}
 \label{tab:ref}
\end{table*}

\subsection{Comparison to Reference Models}

We compare our proposed model against three SOTA reference models: Nakamura's HMM matcher~\cite{nakamura2017performance}, Peter's DualDTWMatcher~\cite{peter23offline} and AutomaticNoteMatcher~\cite{peter2023tismir}.
The first one is implemented in C++ and compiled locally\footnote{downloaded from: \url{https://midialignment.github.io/AlignmentTool_v190813.zip}}, the other's are part of a python package\footnote{downloaded from: \url{https://github.com/sildater/parangonar}}.
The test data consists of five challenging pieces for solo piano in two settings, one default and one with mismatches.
\tabref{tab:ref} details the results.
All values are note match F-scores given in percent, except for the runtime given in seconds.
In the default setting, all three model configurations perform on par with the best model with the best reference model "DualDTWMatcher".
In the mismatch setting, all reference models show (partial) failure.
Note that not all alignments fail, however, no reference model stays consistently above 90 \%.
Our proposed models take a performance loss as well, yet only in the range of 0-7 \% and the F-score stays above 92 \% throughout.
In terms of F-score, no significant difference between TheGlueNote configurations is found.
Despite several forward passes to retrieve local similarity matrices, TheGlueNote configurations also require the lowest runtime.
Unlike for the reference models the runtime does not seem to vary with the complexity of the match to be performed, only with the number of notes and the network size.
The advantageous runtime comparison with the reference models is surprising and to be taken with a grain of salt as implementation details possibly outweigh the merits of each algorithm.

\section{Discussion}\label{sec:discussion}

In this article, we presented TheGlueNote, a note representation model which effectively predicts note matching similarities.
Despite the fundamental role of (note) alignment in several MIR areas, machine learning approaches have seen limited adoption so far --- in contrast with many other areas of MIR, where machine learning models virtually superseded more traditional approaches.
We can only conjecture on the reasons for this absence, however, it seems to us that sequence alignment methods faithfully model a variety of alignment problems and the established methods' correspondingly high performance leaves little room for improvement.
However, this observation does not hold for the question which sequence representations are to be processed by alignment algorithms, a question that is not settled, neither in the symbolic nor in the audio domain.
Feature representations and local metrics abound, and have myriad downstream implications for alignment success or failure which are often hard to predict.

Our approach excels at this point by producing learned representations which leverage non-local information.
The representations play well with DTW-based post-processing, however, end-to-end note matching remains challenging.
Learning representations shifts the problem of edge cases from the modeling stage (or even post-hoc engineering) towards the training data.
Augmenting data with complex mismatches in combination with a model that effectively predicts matches frees us from having to address all possible cases explicitly.
Randomized training mismatches enable the model to learn robust representation for a variety of mismatching sequences.
In practice, this also leads to greater flexibility, as no quantized music, score annotations, or any attributes beyond the basic MIDI notes are required.

Many extensions of our approach are possible.
The hyperparameter and architectural space of plausible representation models open several possibilities for future research.
Furthermore, the data used to train and test the model is specific: solo piano pieces and performances of common practice period music.
This is due to the fact that reference models work on the piano data, and large symbolic piano datasets are available.
Note that this data presents one of the most challenging note alignment scenarios and we expect our core ideas to translate to other symbolic data --- after retraining.
An open question is whether this type of token-based match representation learning can be used in audio or multimodal domains, e.g.~by applying it to discrete audio encodings.
Lastly, the representation learning backbone is trained without any information about the DTW post-processing.
SoftDTW~\cite{cuturi2017soft} approaches appear promising to bridge this gap while keeping sensible alignment constraints in an end-to-end model.
However, we want to stress again that the monotonicity condition of (soft)DTW does not strictly hold in symbolic music even though it has proven an effective heuristic.
Many questions remain open, yet we hope to have shown that representation learning can be integrated successfully and beneficially into note alignment methods.

\section{Reproducibility}\label{sec:reproducibility}
\new{Code and pre-trained checkpoints and public data available at: \url{https://github.com/sildater/thegluenote}}

\section{Acknowledgments}

\new{This work receives funding from the European Research Council (ERC), under the European Union's Horizon 2020 research and innovation programme, grant agreement No.~101019375 (\textit{Whither Music?}). The LIT AI Lab is supported by the Federal State of Upper Austria.}


\bibliography{ISMIRtemplate}

\end{document}